\documentclass{article}
\usepackage[utf8]{inputenc}
\usepackage{amsmath}
\usepackage{graphicx}
\usepackage{enumitem}

\begin{document} 

  \begin{center}

 \Large (1+1) dimensional scalar field theory on q-deformed space

 \end{center}

 \begin{center}
     Poula Tadros\\

     Department of Applied Physics, Aalto University School of Science, FI-00076 Aalto, Finland.\\

     email:poulatadros9@gmail.com
     
 \end{center}




\abstract{    We study scalar field theory in one space and one time dimensions on a q-deformed space with static background. We write the Lagrangian and the equation of motion and solve it to the first order in $q-1$ where $q$ is the deformation parameter of the space.}


\section{Introduction}
Non-commutative geometry was first introduced in string theory in reference [1], where it was shown that the coordinates of the endpoints of strings on D-branes in the presence of a Neveu-Schwartz field are non-commutative. Non-commutative field theories have also been defined, as they can be derived from string theories and have interesting features, as described in references [2] and [3].\\

The introduction of non-commutative spacetime in field theory is motivated by the Heisenberg uncertainty principle in quantum mechanics, which states that at small distance scales, there is a large uncertainty in momentum measurement. This means that energy can reach very high values in a small spatial distance, approaching the Planck scale. However, according to the general theory of relativity, high energy in a small spatial distance creates a black hole, which prevents the position from being fully certain. To reconcile these two phenomena, it is necessary to introduce non-commutativity in spacetime, which implies non locality in the theory. This is explained in references [4] and [5].\\

In this paper we study (1+1) dimensional classical scalar field theory with static spacetime on a q-deformed space, we present both analytical and numerical analysis of the resulting theory. In section 2, we review the some types of non-commutativity on space times and motivate the choice of q-deformation non-commutativity as the subject of the study . In section 3. we study the scalar field theory on q-deformed space time, we write the Lagrangian and deduce the equation of motion, we also truncate the equation of motion to the linear order in $q-1$ and solved the resulting equation. In section 4, we study the numerical solutions of the truncated equation of motion showing that the solutions grow exponentially with $x$ and $t$ meaning that the equation is stiff and there are instabilities in the theory. In section 5, we conclude the study and suggest topics for further research.\\

\section{Types of non-commutativity}

Here, we briefly review three of the most popular types of non-commutativity relations and justify our motivation to use the q-deformation type
\begin{enumerate}

\item Canonical non-commutativity
It is the simplest type which used in physics literature, it was introduced in [6], it is defined by imposing the following commutation relations 
$$[x^{\mu},x^{\nu}]=i\theta^{\mu \nu},$$
where $x^{\mu}$ are the spacetime coordinates and $\theta^{\mu \nu}$ is a constant, anti-symmetric matrix.\\

The idea of canonical non-commutativity involves smearing the structure of space-time in a particular way, regardless of the specific mathematical details of the space. In order to incorporate non-commutative geometry capturing the mathematical structures on the manifold, it is necessary to consider more complex forms of non-commutativity beyond just this basic version.

\item Lie-type non commutativity

In this case the coordinates has a Lie algebra structure i.e. the commutation relations can capture a Lie algebra structures [7]. The commutation relations are given by

$$[x^{\mu}, x^{\nu}]=if^{\mu \nu}_{\rho}x^{\rho},$$
where $f^{\mu \nu}_{\rho}$ are the structure constants of the defined Lie algebra. However, this type is not useful because Lie structures are rigid i.e. any small deformation of a Lie algebra is isomorphic to the original Lie algebra. 

\item q-deformations

A solution to the rigidity problem for Lie algebras is to replace Lie group with a flexible structure called quantum groups [8-10]. The term quantum group used in this context refers to the deformations of the universal enveloping algebra of a given group, these objects have Hopf algebra structures which are flexible structures unlike Lie groups and algebras.\\

The commutation relations are given by

$$x^{\mu} x^{\nu}= \frac{1}{q} R^{\mu \nu}_{\sigma \tau}x^{\sigma}x^{\tau},$$
where $q$ is a parameter and $R^{\mu \nu}_{\sigma \tau}$ is the R-matrix of the quantum group defined on the space.\\

In this space a Lie algebra is replaced by a non-commutative Hopf algebra with deformation parameter q. The resulting space is deformed according to the Lie group on the space and on the parameter q, this is the simplest way to deform a space time while capturing the full algebraic structure of the space.

\end{enumerate}

\section{Lagrangian and the equation of motion}

We begin with the Lagrangian of the scalar field on the commutative manifold then introduce non-commutativity by replacing the derivatives by Jackson derivatives, since the symmetry group is $U(1)$, the deformations of its universal enveloping algebra gives a commutative algebra. Thus, we do not have to worry about defining a product of functions on the new space. The Lagrangian is then

$$\mathcal{L}=\frac{1}{2}\partial^{\mu} \phi \partial_{\mu} \phi -\frac{1}{2}m^2 \phi^2 \rightarrow \mathcal{L}_q=D_{\mu q}\phi D^{\mu}_q \phi -m^2 \phi^2,$$
where $\mu=0,1$ with $x^0=t$ and $x^1=x$.

Assuming the field is defined everywhere and is infinitely differentiable and the deformations are small i.e. $q \approx 1$, we can relate the theory on the non commutative topological space to the theory on the commutative manifold (i.e. transforming the non-commutative theory back to the commutative manifold) using the formulae
 $$D_{xq }(f(x))= \partial_x f+ \sum_{k=1}^{\infty}\frac{(q-1)^k}{(k+1)!}x^k f^{(k+1)}(x) ,$$ where $f^{(k)}$ is the k-th ordinary derivative of f with respect to x.\\
 $$D_{tq }(f(x))= \partial_t f+ \sum_{k=1}^{\infty}\frac{(q-1)^k}{(k+1)!}x^k f^{(k+1)}(x) ,$$ where $f^{[k]}$ is the k-th ordinary derivative of f with respect to t.\\

The resulting Lagrangian on the commutative manifold is
$$\mathcal{L}_q=\frac{1}{2}\partial\phi \partial \phi - \frac{1}{2} m^2 \phi^2 + 2\partial\phi \sum_{k=1}^{\infty}\frac{(q-1)^k}{(k+1)!}x^{ k} \phi^{(k+1)} $$ 
$$ +\sum_{l,m=1}^{\infty}\frac{(q-1)^{(l+m)}}{(m+1)!(l+1)!}x^{k+l} \phi^{(l+1)} \phi^{(m+1)}+ (x\rightarrow t).$$

where $(x\rightarrow t)$ means the same terms but with $x$ replaced by $t$ including in the derivatives.\\

The Lagrangian has an infinite series of derivatives, in this case the Euler-Lagrange equation will be
\begin{equation}
\frac{\partial \mathcal{L}_q}{\partial \phi}+\sum_{k=1}^{\infty}(-1)^k\frac{d^k}{dx^k}(\frac{\partial \mathcal{L}_q}{\partial \phi^{(k)}})+\sum_{k=1}^{\infty}(-1)^k\frac{d^k}{dt^k}(\frac{\partial \mathcal{L}_q}{\partial \phi^{[k]}})=0,
\end{equation}
where $k=2,3,...$.

The Lagrangian is clearly non local as expected from a non-commutative theory.\\

The derivatives of the Lagrangian are given by
$$\frac{\partial \mathcal{L}_q}{\partial \phi}=-m\phi,$$

\begin{equation}
\frac{\partial \mathcal{L}_q}{\partial (\partial \phi)}=\partial_x \phi + 2 \sum_{n=1}^{\infty}\frac{(q-1)^n}{(n+1)!}x^n\phi^{(n+1)}
\end{equation}

$$\rightarrow \frac{d}{dx}(\frac{\partial \mathcal{L}_q}{\partial (\partial_x \phi)})$$
\begin{equation}
= \partial_x \partial_x \phi + 2\sum_{n=1}^{\infty}(\frac{n(q-1)^n}{(n+1)!}x^{n-1}\phi^{(n+1)})+2\sum_{n=1}^{\infty}(\frac{(q-1)^n}{(n+1)!}x^{n}\phi^{(n+2)}),
\end{equation}

$$\frac{\partial \mathcal{L}_q}{\partial (\phi^{(k)})}=\frac{2 (q-1)^{k-1} x^{k-1}}{k!}\sum_{n=0}^{\infty} \frac{(q-1)^n}{(n+1)!}x^n\phi^{(n+1)}$$

$$
\rightarrow  \frac{d}{dx}(\frac{\partial \mathcal{L}_q}{\partial (\phi^{(k)})})
$$
\begin{equation}
=\frac{2(q-1)^{k-1}x^{2k-1}}{k!}\sum_{m,n=0}^{\infty} \binom{k}{m} \frac{(q-1)^n}{(n+1)!} \frac{(n+k+1)!}{(n+2k-m-1)!}x^{n-m} \phi^{(n+k+1)},
\end{equation}
with similar formulae for derivatives with respect to t.\\

Putting all together from (2), (3), (4) in (1) we get

$$
    -\partial_{\mu}\partial^{\mu} \phi -m^2 \phi - 2 \sum_{n=1}^{\infty} \frac{n (q-1)^n}{(n+1)!}x^{n-1} \phi^{(n+1)} - 2 \sum_{n=1}^{\infty} \frac{ (q-1)^n}{(n+1)!}x^{n} \phi^{(n+2)}$$

$$
    + \sum_{k=2}^{\infty} (-1)^k \frac{2 (q-1)^{k-1}x^{2k-1}}{k!}\sum_{n=0}^{\infty}\sum_{m=0}^k \binom{k}{m}\frac{(q-1)^n (n+k-1)!}{(n+1)!(n+2k-m-1)!} x^{n-m} \phi^{(n+k+1)}$$
    \begin{equation}
    + (x \rightarrow t)=0.
\end{equation}

This is a partial differential equation of infinite order with variable coefficients.\\

If we consider only small deformations i.e. $q \approx 1$, then we can only keep terms up to the linear order in $q-1$, the first order equation will be

$$-\partial_{\mu}\partial^{\mu}\phi- m^2 \phi - (q-1)[\phi^{(2)}+x \phi^{(3)} - \frac{x^3}{6}\phi^{(3)} -x^2 \phi^{(3)} - x \phi^{3} + (x \rightarrow t)]=0.$$

This equation is a stiff equation i.e. it is numerically unstable, this may indicate an instability in the theory due to the linear approximation used, but as seen from the full equation of motion the full theory is stable.

The solution is $\phi = F(t)G(x)$ where 
$$F(t)=c_1e^{iAt/\sqrt{q}}+c_2e^{-iAt/\sqrt{q}}+(q-1)\frac{e^{iAt/\sqrt{q}}}{2iA\sqrt{q}}[\frac{iA}{24q}t^4+ (\frac{iA^3}{3}-\frac{1}{12\sqrt{q}}-\frac{i}{8A})t^3$$

$$+(\frac{A+q}{2q}-\frac{A\sqrt{q}}{2}-\frac{i}{8A})t^2+(\frac{i(A+q)}{2A\sqrt{q}}-\frac{iqA}{4}+\frac{\sqrt{q}}{8A^2})t$$
$$+(\frac{A+q}{4A^2}+\frac{q\sqrt{A}}{4}+\frac{iq}{16A^3})]+O((q-1)^2),$$

$$G(x)=c_3e^{ikx/\sqrt{q}}+c_4e^{-ikx/\sqrt{q}}+(q-1)\frac{e^{ikx/\sqrt{q}}}{2ik\sqrt{q}}[\frac{ik}{24q}x^4+ (\frac{ik^3}{3}-\frac{1}{12\sqrt{q}}-\frac{i}{8A})x^3$$

$$+(\frac{k+q}{2q}-\frac{k\sqrt{q}}{2}-\frac{i}{8k})x^2+(\frac{i(k+q)}{2k\sqrt{q}}-\frac{iqk}{4}+\frac{\sqrt{q}}{8k^2})x$$
$$+(\frac{k+q}{4k^2}+\frac{q\sqrt{k}}{4}+\frac{iq}{16k^3})]+O((q-1)^2),$$

where $c_1,c_2,c_3, c_4, A$ are normalisation constants and $k=\pm \sqrt{A^2+m^2}$. When $q=1$, it reduces to the solution to the Klein Gordon equation as expected.\\

\section{Numerical results}
Here, we present numerical solutions to the equation of motion to the first order in $q-1$, we focus on $G(x)$ only since the remaining part is similar. The solutions are exponentially growing in time establishing that the equation of motion was stiff.\\

We set $c_3=c_4=k=1$, $A^2=\frac{1}{2}$ and we plot the solution for different values of the parameter $q$.

\begin{figure}[htbp]
  \centering
 \includegraphics[scale=0.5]{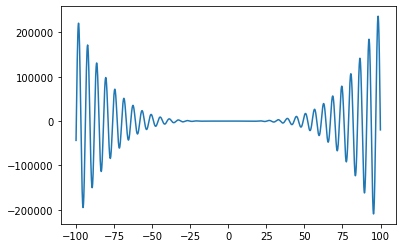}
  \caption{At $q-1=0.1$ the solution grows exponentially with $|x|$. This is a feature of a stiff equation with unstable numerical solution. In the vicinity of $x=0$ it is close to the usual Klein-Gordon solution but as we go further it becomes more and more distant }
  \label{fig:example-image}
\end{figure}

\begin{figure}[htbp]
  \centering
  \includegraphics[scale=0.5]{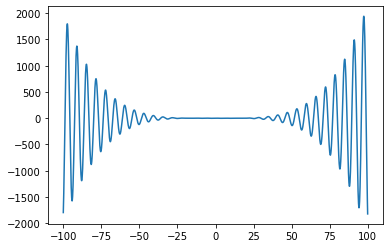}
  \caption{At $q-1=0.001$ the solution still grows exponentially but slower.}
  \label{fig:example-image}
\end{figure}

\begin{figure}[htbp]
  \centering
  \includegraphics[scale=0.5]{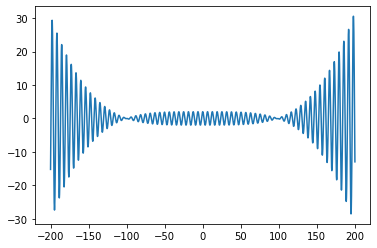}
  \caption{At $q-1=10^{-6}$ the solution resembles the Klein-Gordon solution up to $|x|=50$ then decays for a bit but eventually blows up. }
  \label{fig:example-image}
\end{figure}

\begin{figure}[htbp]
  \centering
  \includegraphics[scale=0.5]{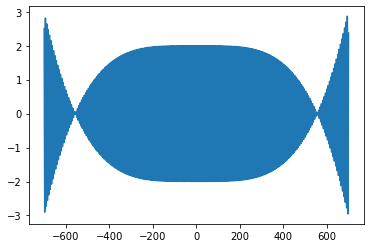}
  \caption{At $q-1=10^{-9}$ the solution has the same behaviour as the previous graph but the decay happens at larger $|x|$, all smaller $q-1$ values follow this pattern.}
  \label{fig:example-image}
\end{figure}

\newpage

The above results shows an instability in the theory leading to divergent solutions to the equations of motion as $x \rightarrow \infty$. To remove the instability we must add infinite terms corresponding to an infinite series of higher derivatives i.e. we have to consider the full theory. However, this approximation gives us an intuition on how the q-deformation affects the space, small q-deformations beside leading to non local effects appear to affect the space irregularly with only small effects locally.

\section{Conclusion and outlook}
In conclusion, we showed that defining a field theory on a q-deformed space leads to an infinite series of higher derivatives in the Lagrangian even with static background. In the case presented the algebra was commutative so no new product of functions is needed. We also demonstrated that any approximation or truncation to the theory will lead to stiff equations of motion resulting from instabilities in the theory.\\
While we made a progress in the field, much more is to be studied, future research in this direction should focus on defining more complicated theories on q-deformed spaces with non-commutative function algebras and with dynamical spacetimes, also to define higher spin fields on such space and study the new symmetries of the theories as well as the types of instabilities arise if the Lagrangian is truncated.

\section*{Acknowledgments}
We would like to thank Dr.Ivan Kolar for the useful discussions on the topic

\section*{References}

\begin{enumerate}[label={[\arabic*]}]

    \item Seiberg, N. and Witten, E. (1999) “String theory and noncommutative geometry,” Journal of High Energy Physics, 1999(09), pp. 032–032.

    \item Szabo, R. (2003) “Quantum field theory on noncommutative spaces,” Physics Reports, 378(4), pp. 207–299.

    \item Sheikh-Jabbari, M.M. (1999) “Super Yang-Mills theory on noncommutative torus from open strings interactions,” Physics Letters B, 450(1-3), pp. 119–125. 

    \item Doplicher, S., Fredenhagen, K. and Roberts, J.E. (1995) “The quantum structure of spacetime at the Planck scale and Quantum Fields,” Communications in Mathematical Physics, 172(1), pp. 187–220.  

    \item Ahluwalia, D.V. (1994) “Quantum measurement, gravitation, and locality,” Physics Letters B, 339(4), pp. 301–303.

    \item C. S. Chu and P. M. Ho, Noncommutative open string and D-brane, Nucl. Phys.
B 550, 151 (1999) [hep-th/9812219].

\item  B. Jurco, S. Schraml, P. Schupp and J. Wess, Enveloping algebra valued gauge
transformations for non-Abelian gauge groups on non-commutative spaces, Eur.
Phys. J. C17, 521 (2000) [hep-th/0006246].

\item Chaichian, M. and Demichev, A.P.  Introduction to quantum groups. Singapore: World Scientific (1996). 

\item Bonneau, P. et al. (2004) “Quantum groups and deformation quantization: Explicit approaches and implicit aspects,” Journal of Mathematical Physics, 45(10), pp. 3703–3741.

\item A. Klimyk and K. Schmudgen, Quantum Groups and Their Representations,
Springer (1997).

\end{enumerate}
\end{document}